\documentclass[showkeys]{revtex4}
\usepackage{graphicx}
\usepackage{dcolumn}
\usepackage{amsmath}
\usepackage{color}
\usepackage{epstopdf}
\usepackage{float}
\usepackage{graphicx}
\usepackage{subfigure}
\usepackage[utf8x]{inputenc}
\definecolor{coolblack}{rgb}{0.0, 0.18, 0.39}
\definecolor{darkred}{rgb}{0.5,0,0}
\definecolor{darkgreen}{rgb}{0,0.5,0}
\definecolor{darkblue}{rgb}{0,0,0.5}
\definecolor{lapislazuli}{rgb}{0.15, 0.38, 0.61}
\definecolor{venetianred}{rgb}{0.78, 0.03, 0.08}
\definecolor{bleudefrance}{rgb}{0.19, 0.55, 0.91}
\definecolor{dogwoodrose}{rgb}{0.84, 0.09, 0.41}
\usepackage[linktocpage,colorlinks]{hyperref}
\hypersetup{colorlinks=true, citecolor=darkgreen, linkcolor=blue,urlcolor = darkblue}
\makeatletter
\def\btt#1{\texttt{\@backslashchar#1}}
\DeclareRobustCommand\bblash{\btt{\@backslashchar}} \makeatother

\RequirePackage{fix-cm}

\begin{document}
\title{Stability analysis of circular orbits around a charged BTZ black hole spacetime in a nonlinear electrodynamics model via Lyapunov exponents%\thanksref{t1}  
}
\author{Shobhit Giri $^{a}$}\email{shobhit6794@gmail.com}
\author{ Hemwati Nandan $^{a,b}$}\email{hnandan@associates.iucaa.in}
\author{Lokesh Kumar Joshi $^{c}$}\email{lokesh.joshe@gmail.com}
\author{Sunil D. Maharaj $^{d}$}\email{maharaj@ukzn.ac.za}
\affiliation{$^{a}$Department of Physics, Gurukula Kangri (Deemed to be University), Haridwar 249 404, Uttarakhand, India}
\affiliation{$^{b}$Center for Space Research, North-West University, Mahikeng 2745, South Africa}
\affiliation{$^{c}$ Department of Applied Science, Faculty of Engineering and Technology, Gurukula Kangri (Deemed to be University), Haridwar 249 404, Uttarakhand, India}
\affiliation{$^{d}$ Astrophysics Research Centre, School of Mathematics, Statastics and Computer Science, University of KwaZulu-Natal, Private Bag X54001, Durban 4000, South Africa }

% The correct dates will be entered by the editor

\begin{abstract}
\noindent  We investigate the existence and stability of both the timelike and null circular orbits for a (2+1) dimensional charged BTZ black hole in Einstein-nonlinear Maxwell gravity with a negative cosmological constant. The stability analysis of orbits are performed to study the possibility of chaos in geodesic motion for a special case of black hole so-called conformally invariant Maxwell spacetime. The computations of both proper time Lyapunov exponent ($\lambda_{p}$) and coordinate time Lyapunov exponent ($\lambda_{c}$) are useful to determine the stability of these circular orbits. We observe the behavior of the ratio $(\lambda_{p}/\lambda_{c})$ as a function of radius of circular orbits for the timelike case in view of different values of charge parameter. However, for the null case, we calculate only the coordinate time Lyapunov exponent ($\lambda_{c}$) as there is no proper time for massless test particles. More specifically, we further analyze the behavior of the ratio of $\lambda_{Null}$ to angular frequency ($\Omega_{c}$), so-called instability exponent as a function of charge ($q$) and parameter related to cosmological constant ($l$) for the particular values of other parameters.\\
% \PACS{PACS code1 \and PACS code2 \and more}
% \subclass{MSC code1 \and MSC code2 \and more}
\end{abstract}
\maketitle
\section{Introduction}
\noindent One of the most robust predictions of Einstein’s general theory of relativity (GR) is the black hole (BH), which attracts theoretical physicists to understand the dynamics of the universe for a long time \cite{Wald1984c,hartle2003gravity,poisson2004relativist,schutz1985first}. One of the Nobel laureates of the current year, Roger Penrose used ingenious mathematical methods to show that BHs are a direct consequence of GR. In 2+1 dimensions, the first exact solution of a vacuum rotating BH with a non-zero cosmological constant arising from collapsing matter was discovered by  Banados, Teitelboim and Zanelli (BTZ) in 1992 \cite{banados1992black}.  The foundation of the classical and quantum gravity aspects is a consequence of the existence of three-dimensional spacetime in general relativity. In order to provide a better understanding of gravitational interactions in low-dimensional spacetimes, the BTZ BHs are the greatest achievements \cite{banados1993geometry,soroushfar2016study,hendi2012asymptotic,hendi2020simulation,jayawiguna2020three,hendi2014exact}. The BTZ BH is asymptotically anti-de-Sitter (AdS) and has no curvature singularity at the origin \cite{banados1992black,hendi2020simulation,acena2020circular,dias2002magnetic,dehghani2004horizonless}. This is the fundamental difference between Schwarzschild and Kerr BH which are asymptotically flat and having a ring singularity at origin. These objects are still BHs having  both the inner horizon and the outer horizon, ergopshere, static limit regions. \\
It is indeed interesting to perform the stability analysis of circular orbits for the special case of a charged BTZ BH solution obtained from the Einstein–nonlinear Maxwell equations with a negative cosmological constant in presence of a matter source with power Maxwell invariant \cite{hassaine2008higher}. The geometry of such a BH solution has emerged out as topic of crucial interest especially due to non-linearity of the system as compared to the usual BTZ BH case \cite{hassaine2008higher,maeda2009lovelock,hendi2009magnetic,hendi2011charged}. Since the charged BTZ BH has no constant curvature, it has not been thoroughly investigated in comparison with the neutral BHs  \cite{jayawiguna2020three,garfinkle1992erratum}. The metric expression of the charged BTZ BH consists of a logarithmic function which is an obstacle to understand the geometrical properties and makes the analytic investigation difficult. Therefore, we intend to investigate the stability of geodesics for the  well-known conformally invariant Maxwell solution derived for the special case of nonlinear parameter $s=3/4$ \cite{soroushfar2016study}. The study of geodesics in a (2+1)–dimensional charged BTZ BH conveying important information on the spacetime geometry has been investigated \cite{soroushfar2016study}. Tang.et.al  \cite{tang2017thermodynamical} have widely studied the thermodynamical and dynamical properties of charged BTZ BHs. The Quasi-Normal Modes (QNMs) under massless scalar field perturbations and the thermodynamics of linearly charged BTZ BHs in massive gravity around the anti-de Sitter (AdS) spacetime have been analyzed by Prasia.et.al \cite{prasia2017quasinormal}. \\
\noindent The nonlinearity of GR is a direct consequence of the instability of circular orbits around  any BH spacetime. In the context of unstable circular orbits, the Lyapunov exponent is a  good indicator of instability and their instability is quantified by a positive Lyapunov exponent \cite{cornish2003lyapunov,cornish2001chaos}. Circular orbits around any BH spacetime can be chaotic if there exists instability in geodesics under perturbation like the spin of BH or spin of the test particle \cite{hilborn2000chaos,Suzuki1997}. \\ 
The existence and stability of circular geodesics around static and spherically symmetric (SSS) spacetimes using the Lyapunov exponent has been extensively studied in detail by many authors \cite{pradhan2016stability,cardoso2009geodesic,pradhan2012isco,pradhan2013lyapunov}. The main motivation of this work is to investigate stability of circular orbits around AdS spacetime in terms of Lyapunov exponents. \\
\noindent The paper is organized as follows. In Section 1.1, we discuss the Lyapunov exponent in order to establish a relation with radial effective potential. In Section 2, we briefly introduce charged BTZ BH solutions and extend them further to investigate stability of timelike and null circular geodesics in Section 2.1 and Section 2.2 respectively. Finally, we conclude the main results and discuss in Section 3.

\subsection{Lyapunov exponents}
The stability of circular orbits (timelike and null) needs careful attention to check whether they are chaotic or not. For this purpose, the Lyapunov exponent in a classical phase space is an important tool to determine the stability of such orbits. The average rate of convergence or divergence of nearby trajectories in any dynamical system is referred as the Lyapunov exponent \cite{sano1985measurement,skokos2010lyapunov,cornish2003lyapunov}. It was shown many years ago that circular geodesics (timelike and null) are needed to investigate whether circular orbits are stable or not. The positive and negative values of Lyapunov exponents indicates divergence and convergence of two nearby geodesics respectively.
We begin with an observed trajectory denoted by $X_{i}$, which is solution of the equations of motion given as \cite{sano1985measurement}

\begin{equation}
	\frac{dX_{i}}{dt}=F_{i}(X_{j})\label{eq1}.
\end{equation}
To study the stability of a given orbit linearizing the Eq.(\ref{eq1}) about that orbit $X_{i}(t)$ we require,
\begin{equation}
	\frac{d\partial X_{i}(t)}{dt}=K_{ij}(t)\partial X_{j}(t)\label{eq2},
\end{equation}
\noindent where, the linear stability matrix $K_{ij}(t)$ is defined as
\begin{equation}K_{ij}(t)=\frac{\partial F_{i}}{\partial X_{j}}\label{eq3}.
\end{equation}
The linearized equation (\ref{eq2}) yields
\begin{equation}
	\partial X_{i}(t)= L_{ij}(t) \partial X_{j}(0)\label{eq4},
\end{equation}
where, $L_{ij}(t)$ is the evolution matrix or operator. The eigenvalues $\lambda_{i}$ of the evolution matrix are known as the “Lyapunov exponents” and the largest of these eigenvalues is known as the principal Lyapunov exponent \cite{cardoso2009geodesic}
\begin{equation}
	\lambda_{0}= \lim_{t\to\infty}\frac{1}{t} ln \left(\frac{ L_{jj}(t)}{ L_{jj}(0)}\right)\label{eq5}.
\end{equation}
\noindent However, the Lyapunov exponent (i.e. proper time Lyapunov exponent $\lambda_{p}$ and coordinate time Lyapunov exponent $\lambda_{c}$) and geodesics around any BH spacetime having radial effective potential, $V_{r}=\dot{r}^{2}$ are correlated by the well established relationship defined as \cite{cardoso2009geodesic,pradhan2016stability},

\begin{equation}
	\lambda_{p}=\pm\sqrt{\frac{V_{r}^{''}}{2}}\label{eq6},
\end{equation}
\begin{equation}
	\lambda_{c}=\pm\sqrt{\frac{V_{r}^{''}}{2 \dot{t}^{2}}}\label{eq7}.
\end{equation}
\noindent Throughout this work, we consider only positive Lyapunov exponent. The circular orbits are unstable corresponding to real values of the Lyapunov exponent $\lambda_{p}$ or $\lambda_{c}$ i.e. for $V_{r}^{''}>0$  whereas, the stable circular orbits correspond to imaginary nature of $\lambda_{p}$ or $\lambda_{c}$ i.e. for $V_{r}^{''}<0$. The circular orbits are called marginally stable when $\lambda_{p}$ or $\lambda_{c}$ vanish i.e. for $V_{r}^{''}=0$  \cite{pradhan2016stability}. \\
Pretorius and Khurana introduced a quantity the so-called the critical exponent ($\gamma$) which refers to a quantitative measure of instability for circular geodesics \cite{pretorius2007black}. This quantity is characterized  by a typical orbital timescale $T_{\Omega}=\frac{2\pi}{\Omega}$ and Lyapunov time scale or instability time scale $T_{\lambda}=\frac{1}{\lambda}$ by an interesting relation \cite{cardoso2009geodesic,pradhan2016stability},

\begin{equation}
	\gamma= \frac{T_{\lambda}}{T_{\Omega}}= \frac{\Omega}{2\pi \lambda}\label{criticalex}~.
\end{equation}
Therefore, the critical exponent $\gamma$ and second order derivative of the radial 
effective potential for proper time and coordinate time i.e. $V_{r}^{''}$ are correlated by

\begin{equation}
	\gamma_{p}= \frac{\Omega}{2\pi \lambda_{p}}=\frac{1}{2\pi}\sqrt{\frac{2\Omega^{2}}{V_{r}^{''}}} \thickspace, ~~~~
	\gamma_{c}= \frac{\Omega}{2\pi \lambda_{c}}=\frac{1}{2\pi}\sqrt{\frac{2 \Omega^{2}\dot{t}^{2}}{V_{r}^{''}}}
	\thickspace.\label{cri expo}
\end{equation}
where $\Omega$ is the angular frequency or orbital angular velocity of the test particle.

\section{ The charged BTZ BH spacetime in a nonlinear electrodynamics model }
Firstly, we review briefly the charged BTZ BH as a solution of the (2+1) dimensional Einstein-nonlinear Maxwell gravity in presence of the nonlinear electrodynamics source with a negative cosmological constant \cite{soroushfar2016study}. The lower (2+1) dimensional  BTZ BH is quite easy to imagine in comparison with familiar (3+1) dimensional BHs like Schwarzschild, Kerr, etc. One can obtain such BH spacetime as a solution of Einstein field equations in (2+1) dimensions with the nonlinear electrodynamic source which is represented by the Maxwell action in (2+1) dimensions as follows \cite{hassaine2008higher}

\begin{equation}
	I_{M}=-\frac{1}{16\pi G}\int \sqrt{-g}(k F)^{s} dx^{3}, \label{actionm}
\end{equation}
where, $G$ stands for gravitational constant,  $k$ is a constant. Here the Maxwell invariant is given by $F=F_{\mu\nu}F^{\mu\nu}$, where, $F_{\mu\nu}=\partial_{\mu}A_{\nu}-\partial_{\nu}A_{\mu}$, refers to the electromagnetic field tensor and $A_{\mu} $ is the gauge potential (or electromagnetic field). The arbitrary positive nonlinear parameter $s$ is restricted to $s>1/2$ for a asymptotically well defined electric field.\\
\noindent The Einstein-nonlinear Maxwell action in presence of cosmological constant for BH in (2+1) spacetime dimensions is given by \cite{hendi2011charged,hassaine2008higher,soroushfar2016study},

\begin{equation}
	I=\frac{1}{16\pi G}\int \sqrt{-g}\left(R-2\Lambda+ (k F)^{s}\right) dx^{3}.\label{action}
\end{equation}
Here, $R$ denotes the Ricci scalar and $\Lambda$ is cosmological constant related to parameter $l$ by $\Lambda=-1/l^{2}$.
It is worth to notice that in absence of Maxwell invariant term in the above action, one can obtain the usual BTZ BH spacetime \cite{banados1992black}.
The equations of gravitational and electromagnetic fields are obtained by varying action $I$ with metric tensor $g_{\mu\nu}$ and electromagnetic field $A_{\mu}$ as follows

\begin{equation}
	G_{\mu\nu}-\Lambda g_{\mu\nu}= T_{\mu\nu},\label{efe}
\end{equation}

\begin{equation}
	\partial_{\mu} \left(\sqrt{-g}F^{\mu\nu}(kF)^{s-1}\right)=0.
\end{equation}

\noindent The energy–momentum tensor in the presence of nonlinear electromagnetic field given by

\begin{equation}
	T_{\mu\nu}= 2\left[skF_{\mu\rho}F^{\rho}_{\nu}(kF)^{s-1}-\frac{1}{4}g_{\mu\nu}(kF)^{s}\right], \label{emt}.
\end{equation}
Specifically, one can deduce the field equations of BH spacetime in Einstein-Maxwell gravity at $s=k=-1$. However in context of nonlinear electrodynamics model, for a well defined electric field with  nonlinearity parameter $s>1/2$, the (2+1) dimensional line element of charged BTZ BH is described by \cite{soroushfar2016study,hendi2012asymptotic,hendi2014exact}

\begin{equation}
	dS^{2}=-f(r) dt^{2}+\frac{dr^{2}}{f(r)}+r^{2}d\phi^{2},
\end{equation}
where the components of Eq.(\ref{efe}) will lead to the metric function $f(r)$ as follows \cite{soroushfar2016study,hendi2014exact},

\begin{equation}
	f(r)= \frac{r^{2}}{l^{2}}-m+ \newcommand{\twopartdef}
	
	\left\{
	\begin{array}{ll}
		{2 q^{2}ln\left(\frac{r}{l}\right)} & \mbox{for } s=1, 
		\\
		
		{\frac{	(2s-1)^{2}\left(\frac{8q^{2}(s-1)^2}{(2s-1)^2}\right)^{s}}{2(s-1)}r^{\frac{2(s-1)}{2s-1}}} & \mbox{otherwise. }
	\end{array}
	\right.
\end{equation}
The metric function is dependent on the radial coordinate and other BH
parameters, such as the mass, charge and cosmological parameter. In this metric, $m$ is the integration constant proportional to the mass of BH ($M$) and $q$ is electric charge of the BH.\\ 
In this article, we will discuss the geodesic stability for the special case $s=3/4$,  for which we obtain a well known spacetime so-called conformally invariant Maxwell spacetime \cite{hendi2014exact}.\\
\noindent The metric function for the limit $s=3/4$ reduces to

\begin{equation}
	f(r)=\frac{r^{2}}{l^{2}}-m-\frac{(2q^{2})^{3/4}}{2r}.
\end{equation}
Replacing $(2q^{2})^{3/4}$ with $K$, the spacetime metric reads as

\begin{equation}
	dS^{2}=-\left(\frac{r^{2}}{l^{2}}-m-\frac{K}{2r}\right) dt^{2}+\frac{dr^{2}}{\left(\frac{r^{2}}{l^{2}}-m-\frac{K}{2r}\right)}+r^{2}d\phi^{2}.
\end{equation}

\noindent The Lagrangian for describing the  geodesic motion in the above mentioned spacetime can be written as

\begin{equation}
	\begin{aligned}
		\mathcal{L}= \frac{1}{2}\sum g_{\mu\nu}\dot{x^{\mu}}\dot{x^{\nu}}=\frac{1}{2}\left[-\left(\frac{r^{2}}{l^{2}}-m-\frac{K}{2r}\right) \dot{t}^{2}+\frac{\dot{r}^{2}}{\left(\frac{r^{2}}{l^{2}}-m-\frac{K}{2r}\right)}+r^{2}\dot{\phi}^{2}\right].\end{aligned}
\end{equation}
Here and throughout the paper, an over dot ($.$) represents the differentiation with respect to proper time ($\tau$).

\noindent The metric is independent of coordinates $t$ and $\phi$ corresponding to which two Killing vectors exist. Using the well known Euler–Lagrange equations of motion, the two constants of motion for particles are obtained as 

\begin{equation}
	P_{t}= -f(r) \dot{t}= -\left(\frac{r^{2}}{l^{2}}-m-\frac{K}{2r}\right)\dot{t}=-E ~,~~~~~ P_{\phi}= r^{2}\dot{\phi}=L \label{momentum},
\end{equation}
and radial component of four-momentum of particle is

\begin{equation}
	P_{r} = \frac{\dot{r}}{f(r)}=\frac{\dot{r}}{\left(\frac{r^{2}}{l^{2}}-m-\frac{K}{2r}\right)},
\end{equation}
where, $E$ and $L$ are the energy and angular momentum receptively.

\noindent The Hamiltonian for the motion of particle around the spacetime is then given by

\begin{equation*}
	2 \mathcal{H} = 2\left( P_{t} \dot{t} +P_{r}\dot{r}+ P_{\phi}\dot{\phi}- \mathcal{L} \right)
\end{equation*}
\begin{equation*}
	~~~~~~~~~~~~~~~~~~~~~~~~~~~~~~~~~~~~~~~=-\left(\frac{r^{2}}{l^{2}}-m-\frac{K}{2r}\right) \dot{t}^{2}+\frac{\dot{r}^{2}}{\left(\frac{r^{2}}{l^{2}}-m-\frac{K}{2r}\right)}+r^{2}\dot{\phi}^{2}
\end{equation*}

\begin{equation}
	~~~~~~~~~~~~~~~~~~=-E\dot{t}+L\dot{\phi}+\frac{\dot{r}^{2}}{\left(\frac{r^{2}}{l^{2}}-m-\frac{K}{2r}\right)}=\epsilon,\label{hamilto}
\end{equation}
with, $\epsilon=0$ and $-1$ correspond to null and timelike geodesics respectively.
Now, by substituting $\dot{t}$ and $\dot{\phi}$ from Eq.(\ref{momentum}) into Eq.(\ref{hamilto}), we obtain

\begin{equation}
	\dot{r}^{2}= E^{2}- \left(\frac{r^{2}}{l^{2}}-m-\frac{K}{2r}\right) \left(\frac{L^{2}}{r^{2}}-\epsilon \right),\label{radial}
\end{equation}
which is referred to the radial equation of motion for charged BTZ BH. The effective potential for radial motion is defined as $V_{r}=\dot{r}^{2}$.

\subsection{Lyapunov exponents and stability of timelike circular orbits}
The radial equation (\ref{radial}) for timelike geodesics with $\epsilon=-1$ is

\begin{equation}
	V_{r}=\dot{r}^{2}= E^{2}- \left(\frac{r^{2}}{l^{2}}-m-\frac{K}{2r}\right) \left(1+\frac{L^{2}}{r^{2}} \right) \label{vr}.
\end{equation}

\noindent To illustrate the stability analysis of circular geodesics, we follow the  method of computing the Lyapunov exponents as described by Cardoso at al. \cite{cardoso2009geodesic}. In order to restrict the motion to circular geodesics, the condition $V_{r}=V_{r}^{'}=0$ must be satisfied, where  the prime ($'$) denotes derivatives with respect to $r$.

\noindent So, Eq.(\ref{vr}) with the condition at a radius of circular orbit $r=r_{0}$ i.e. $V_{r}= V_{r}^{'}=0$, yields the energy and angular momentum of the particle respectively as

\begin{equation}
	E_{0}^{2}=-\frac{\left(Kl^{2}+2l^{2}mr_{0}-2r_{0}^{3}\right)^{2}}{l^{4} r_{0}\left(3K+4 m r_{0}\right)}~,
\end{equation}

\begin{equation}
	L_{0}^{2}= -\frac{r_{0}^{2}\left(K l^{2}+4 r_{0}^{3}\right)}{l^{2}\left(3K+4m r_{0}\right)}~.
\end{equation}

\noindent Since, the energy of the particle must be real and finite, one requires

\begin{equation} 3K+4m r_{0} <0 .\label{con1}\end{equation}
Also, the angular momentum is necessarily to be positive (i.e. $L_{0}^{2}>0$), so one also requires

\begin{equation} Kl^{2}+4r_{0}^{3} >0 .\label{con2}\end{equation}

\noindent Hence, the above two conditions imply that the timelike circular geodesics do not exist for $K=0$, whereas, for $K>0$, the condition (\ref{con1}) leads to

\begin{equation} r_{0} > r_{E}=\frac{3K}{4m},\end{equation}
and from condition (\ref{con2}), we obtain

\begin{equation} r_{0} < r_{L}=\left(\frac{Kl^{2}}{4}\right)^{1/3},\end{equation}
where $r_{E}$ and $r_{L}$ represent energy and angular momentum conditions.

\noindent Therefore, in this case we found that $r_{E}<r_{L}$, which confirms the non-existence of circular geodesics for $K \geq 0$.\\
\noindent For $K<0$, the conditions (\ref{con1}) \& (\ref{con2}) provide

\begin{equation} r_{0} < r_{E}=\frac{3K}{4m} ,~~~~~~~~r_{0} > r_{L}=\left(\frac{Kl^{2}}{4}\right)^{1/3},\end{equation}
\noindent i.e. $r_{E}>r_{L}$ , which implies the existence of circular geodesics for $K<0$.\\

\noindent The angular frequency or orbital angular velocity for the timelike circular orbit is given by

\begin{equation}
	\Omega_{0} = \frac{\dot{\phi}}{\dot{t}}= \sqrt{\frac{\left(Kl^{2}+4r_{0}^{3}\right)}{4l^{2}r_{0}^{3}}}.
\end{equation}

\noindent Now, the second derivative of the radial effective potential with respect to $r$ is 

\begin{equation}
	V_{r}^{''} =-\left[\frac{6\left(Kl^{2}+4r_{0}^{3}\right)\left(K+mr_{0}\right) }{l^{2}r_{0}^{3}\left(3K+4m r_{0}\right)}+\frac{\left(2r_{0}^{3}-Kl^{2}\right)}{l^{2}r_{0}^{3}}\right].
\end{equation}

\noindent We further calculated the proper time Lyapunov exponent for timelike circular geodesics as

\begin{equation}
	\lambda_{p}= \sqrt{\frac{V_{r}^{''}}{2}}= \sqrt{-\left[\frac{3\left(Kl^{2}+4r_{0}^{3}\right)\left(K+mr_{0}\right) }{l^{2}r_{0}^{3}\left(3K+4m r_{0}\right)}+\frac{\left(2r_{0}^{3}-Kl^{2}\right)}{2l^{2}r_{0}^{3}}\right]},
\end{equation}
and the coordinate time or principal Lyapunov exponent is expressed as

\begin{equation}
	\lambda_{c}= \sqrt{\frac{V_{r}^{''}}{2\dot{t}^{2}}}= \sqrt{\frac{\left(3K+4mr_{0}\right)}{4r_{0}}\left[\frac{3\left(Kl^{2}+4r_{0}^{3}\right)\left(K+mr_{0}\right) }{l^{2}r_{0}^{3}\left(3K+4m r_{0}\right)}+\frac{\left(2r_{0}^{3}-Kl^{2}\right)}{2l^{2}r_{0}^{3}}\right]}.
\end{equation}

\begin{figure}[H]
	\centering
	\subfigure[]{\includegraphics[width=6cm,height=6.5cm]{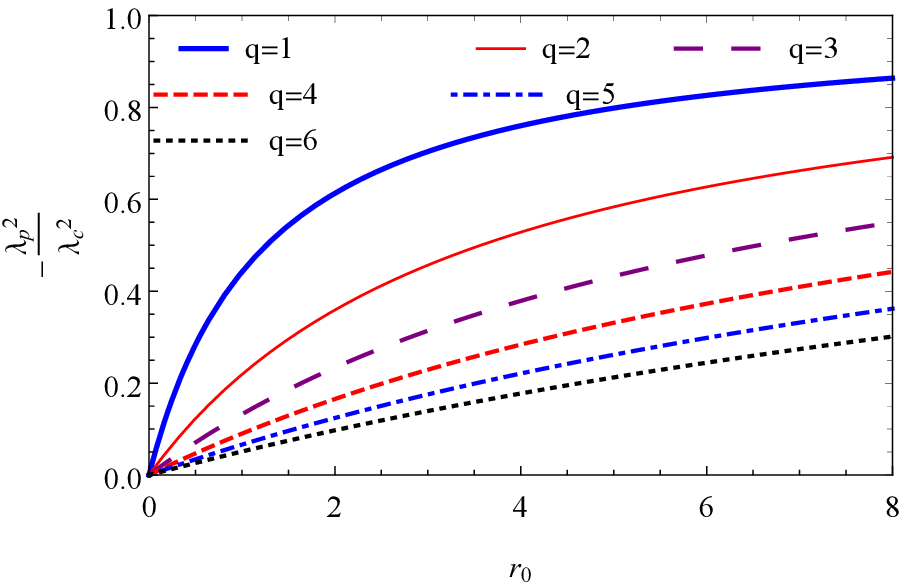}\label{LEratioa}}
	\subfigure[]{\includegraphics[width=6cm,height=6.5cm]{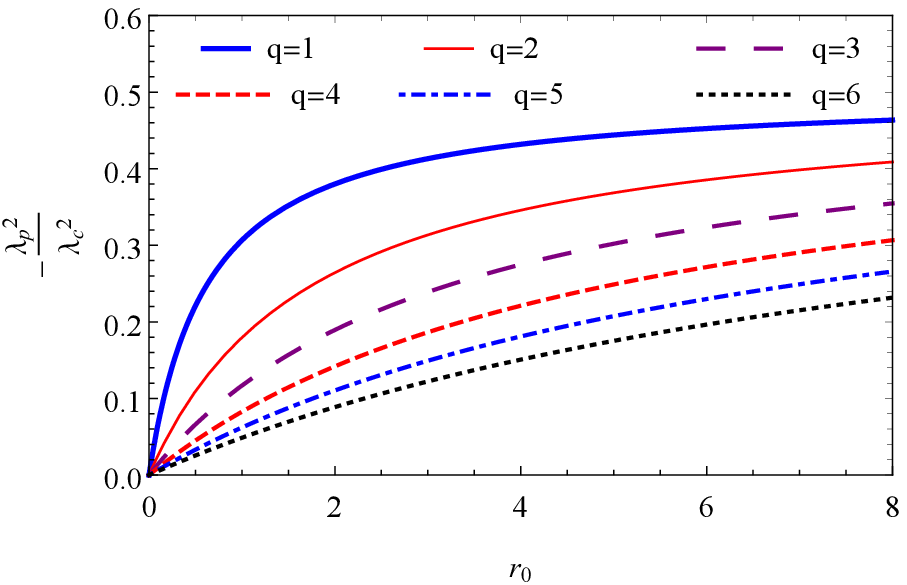}\label{LEratiob}}
	\caption {Behavior of ratio  $-\lambda_{p}^{2}/\lambda_{c}^{2}$ as a function of $r_{0}$ with different values of charge parameter ($q = 1,2,3,4,5,6$) and for fixed value of mass (a) $m=1$ (left panel) (b) $m=2 $ (right panel) .}\label{LEratio}
\end{figure}
\noindent For further descriptions, it is convenient to define the bracketed term in the above expressions as

\begin{equation}
	\Sigma =\left[\frac{3\left(Kl^{2}+4r_{0}^{3}\right)\left(K+mr_{0}\right) }{l^{2}r_{0}^{3}\left(3K+4m r_{0}\right)}+\frac{\left(2r_{0}^{3}-Kl^{2}\right)}{2l^{2}r_{0}^{3}}\right].
\end{equation}

\noindent The unstable circular orbits are described by $V_{r}^{''}>0$, such that Lyapunov exponents must be real, for which $\Sigma<0$. The timelike circular geodesics are stable when Lyapunov exponents are imaginary i.e. $V_{r}^{''}<0$ , for which  $\Sigma>0$. The instability of geodesics can be quantitatively characterized by computing the timescale associated with it. This quantity is known as the critical exponent $(\gamma)$ associated with unstable circular geodesics.\\

\noindent Eq.(\ref{cri expo}) gives the critical exponent corresponding to proper time and coordinate time respectively as follows

\begin{equation}
	\gamma_{p}= \frac{\Omega_{0}}{2\pi \lambda_{p}}=\frac{1}{2\pi}\sqrt{-\frac{\left(Kl^{2}+4r_{0}^{3}\right)}{4l^{2}r_{0}^{3}~ \Sigma}},
\end{equation}
\\
\begin{equation}
	\gamma_{c}= \frac{\Omega_{0}}{2\pi \lambda_{c}}=\frac{1}{2\pi}\sqrt{\frac{\left(Kl^{2}+4r_{0}^{3}\right)}{l^{2}r_{0}^{2}\left(3K+4mr_{0}\right) \Sigma}}.
\end{equation}

\noindent Therefore, for any unstable circular orbit, $\Sigma<0$ must be satisfied so that the Lyapunov timescale is shorter than the orbital timescale ($T_{\lambda}<T_{\Omega}$) which implies observational relevance of instability of circular orbits.

\noindent In fact, it is interesting to visualize the ratio of Lyapunov exponents ($\lambda_{p}/\lambda_{c}$) with respect to radius of circular orbit ($r_{0}$) for different values of electric charge $q$ of BH. The ratio is deduced as

\begin{equation}
	\frac{\lambda_{p}}{\lambda_{c}}=\sqrt{-\frac{4 r_{0}}{\left(3K+4m r_{0}\right)}}.
\end{equation}
The behavior of the ratio $-\lambda_{p}^{2}/\lambda_{c}^{2}$ with $r_{0}$ for different values of charge parameter $q$ and for fixed values of mass $m$, are presented in \figurename{\ref{LEratio}}. One can clearly notice that the Lyapunov exponents ratio varies for one orbit to another orbit for different values of the charge parameter.

\subsection{Lyapunov exponents and stability of null circular orbits}
In this section, we illustrate the motion of massless test particles (like photons) around charged BTZ BH  to discuss stability of null circular geodesics. From Eq.(\ref{radial}) with limit $\epsilon=0$, we obtain the radial potential for the present case as follows

\begin{equation}
	V_{r}=\dot{r}^{2}= E^{2}- \left(\frac{r^{2}}{l^{2}}-m-\frac{K}{2r}\right) \left(\frac{L^{2}}{r^{2}} \right).
\end{equation}
We further derive a relation between energy and angular momentum defined at the radius $r=r_{c}$ of a null circular geodesic with  $V_{r}=0$ as

\begin{equation}
	\frac{E_{c}}{L_{c}} = \pm \sqrt{\frac{\left(2r_{c}^{3}-2 m l^{2} r_{c}-Kl^{2}\right)}{2l^{2}r_{c}^{3}}}\thickspace.\label{eq42}
\end{equation}

\noindent Now by considering the condition $V_{r}^{'}=0$, one can obtain the position of the null circular geodesics which are located at
\begin{align}
	r_{c}=\frac{-3K}{4m}\thickspace,
\end{align}
called the radius of null circular orbit.\\

\noindent Hence, it is observed that the circular geodesics do not exist for $K\geq0$. The relation (\ref{eq42}) between $E_{c}$ and $L_{c}$ exists only for $K<0 $ which implies the existence of null circular geodesics.

\noindent However, the impact parameter associated with null circular geodesics is given by,

\begin{equation}
	D_{c}= \frac{L_{c}}{E_{c}}= \sqrt{\frac{2l^{2}r_{c}^{3}}{\left(2r_{c}^{3}-2 m l^{2} r_{c}-Kl^{2}\right)}}\thickspace.
\end{equation}

\noindent The angular frequency $\Omega_{c}$ at $r=r_{c}$, which is an important quantity for the analysis of null circular geodesics is calculated as

\begin{equation}
	\Omega_{c}= \frac{\dot{\phi}}{\dot{t}} = \sqrt{\frac{\left(2r_{c}^{3}-2 m l^{2} r_{c}-Kl^{2}\right)}{2l^{2}r_{c}^{3}}}=\frac{1}{D_{c}}\thickspace.
\end{equation}
Thus, one can conclude that the angular frequency corresponding to null circular geodesics is inversely proportional to the impact parameter associated with it.

\noindent The second derivative of the radial potential comes out to be

\begin{equation}
	V_{r}^{''}= \frac{6 L_{c}^{2}\left(K+m r_{c}\right)}{r_{c}^{5}}.
\end{equation}

\noindent We finally obtained the coordinate time Lyapunov exponent of null circular geodesics for charged BTZ BH in the form 

\begin{equation}
	\lambda_{Null}=\sqrt{\frac{V_{r}^{''}}{2\dot{t}^{2}}}=\sqrt{\frac{3}{2}\frac{\left(K+mr_{c}\right)}{l^{2}r_{c}^{4}}}\thickspace.
\end{equation} 
\noindent Hence, for the radius of the circular orbit $r_{c}=-3K/4m $, the Lyapunov exponent ($\lambda_{Null}$) is real and finite which indicates existence of unstable null circular geodesics.\\

\noindent Now, the critical exponent corresponding to null circular geodesics of charged BTZ BH is given by,

\begin{equation}
	\gamma_{Null} = \frac{T_{\lambda}}{T_{\Omega}}= \frac{\Omega_{c}}{2\pi \lambda_{Null}}=\frac{1}{2\pi}\sqrt{\frac{r_{c}\left(2r_{c}^{3}-2 m l^{2} r_{c}-Kl^{2}\right)}{3\left(K+m r_{c}\right)}}.
\end{equation}
\noindent In order to measure how fast instability would be noticed, one can compare the instability timescale (or Lyapunov timescale) to the unstable circular geodesics ($T_{\lambda}$) with the orbital timescale ($T_{\Omega}$). For the present case,  $T_{\lambda}<T_{\Omega}$ as $\lambda_{Null}$ is a real value, which is of significance for instability of geodesics. \\

\noindent The Lyapunov exponent and angular frequency corresponding to the null geodesic are crucial to understand the instability of unstable circular orbits. One can therefore represent the instability exponent ($\lambda_{Null}/\Omega_{c}$) in the following form

\begin{equation}
	\frac{\lambda_{Null}}{\Omega_{c}}= \sqrt{\frac{3\left(K+m r_{c}\right)}{r_{c}\left(2r_{c}^{3}-2 m l^{2} r_{c}-Kl^{2}\right)}}.
\end{equation}

\begin{figure}[H]
	\centering
	\subfigure[]{\includegraphics[width=6cm,height=6.5cm]{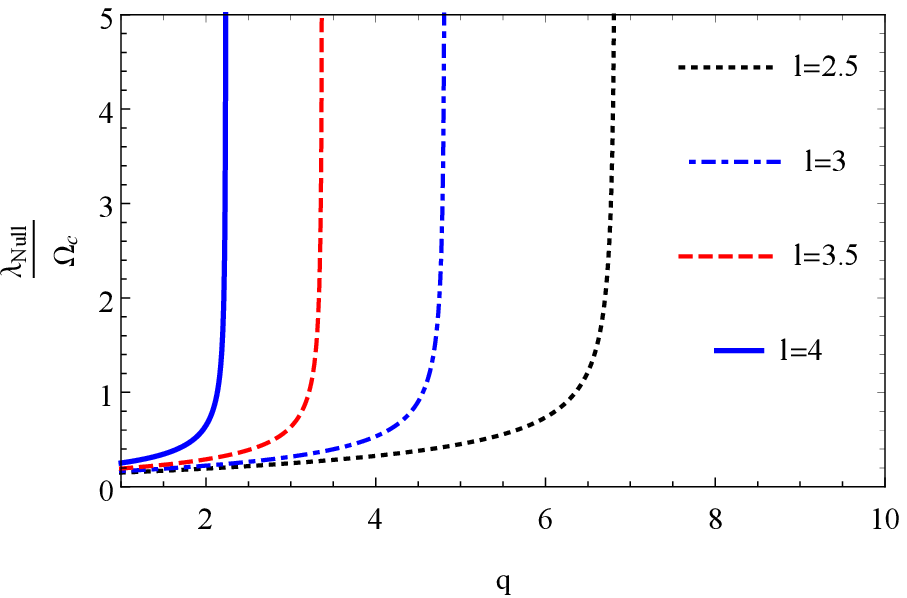}\label{LbyOa}}
	\subfigure[]{\includegraphics[width=6cm,height=6.5cm]{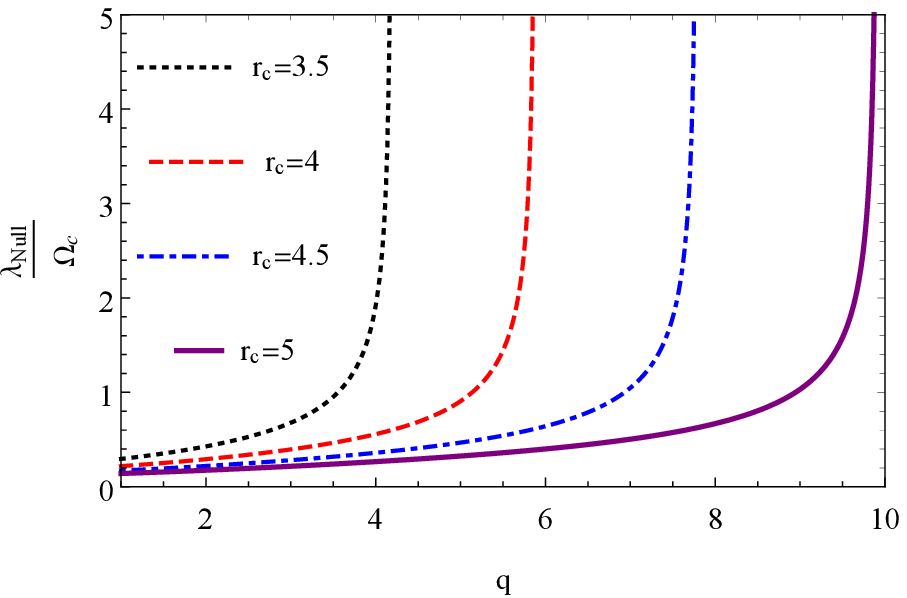}\label{LbyOb}}
	\caption { Behavior of  $\lambda_{Null}/\Omega_{c}$ as a function of charge $q$: (a) for  different values of the cosmological parameter $l$ with $m=1$ and a particular value of radius $r_{c}=5$, (b)  for  different values of radius $r_{c}$ with $m=1$ and a particular value of parameter $l=2$.}
\end{figure}

\begin{figure}[H]
	\centering
	\subfigure[]{\includegraphics[width=6cm,height=6.5cm]{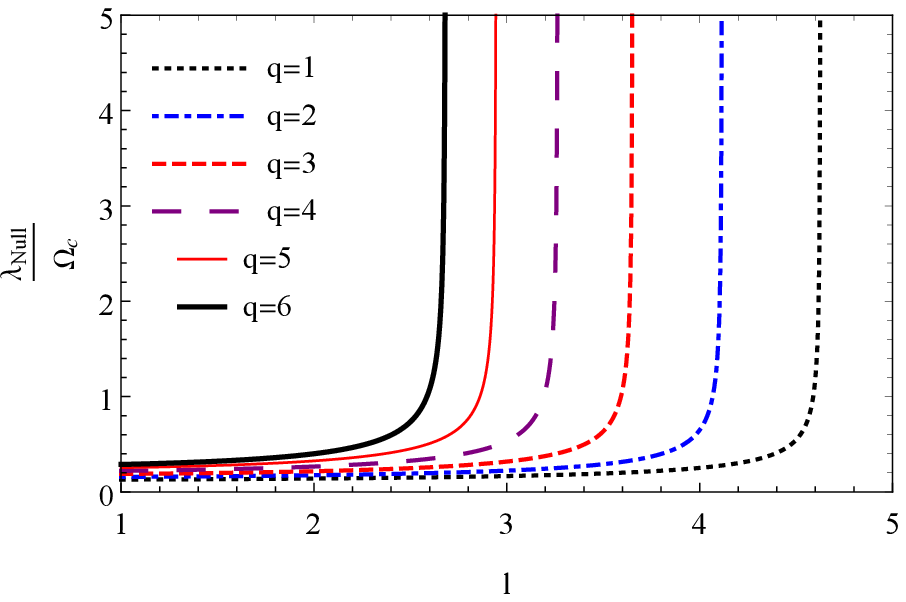}\label{LbyOc}}
	\subfigure[]{\includegraphics[width=6cm,height=6.5cm]{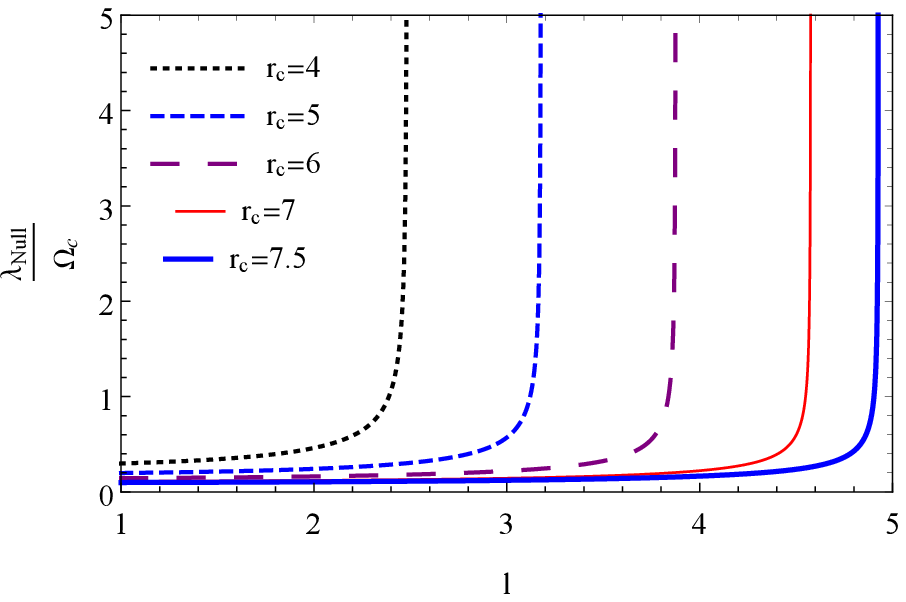}\label{LbyOd}}
	\caption {Behavior of  $\lambda_{Null}/\Omega_{c}$ as a function of cosmological parameter $l$: (a) for  different values of charge $q$ with $m=1$ and a particular value of radius $r_{c}=5$, (b)  for  different values of radius of orbit $r_{c}$  with $m=2$ and a particular value of charge $q=2$.}
\end{figure}

\noindent  The variation of the instability exponent, $\lambda_{Null}/\Omega_{c}$ as a function of charge $q$ for different values of cosmological parameter $l$, by making the value other parameters fixed is presented in \figurename{\ref{LbyOa}}. It can be observed that for higher values of $l$, maximum instability shifts towards lower values of charge $q$. On the other hand, \figurename{\ref{LbyOb}} shows the variation of the same for various values of $r_{c}$ which indicates that for higher values of $r_{c}$, the maximum instability shifts towards large values of $q$. \\
However, it is also interesting to observe the variation of instability exponent with cosmological parameter $l$ for different values of $q$ which can be visualized in \figurename{\ref{LbyOc}}. One can observe that the maximum instability shifts towards lower values of $l$ for an increase in the values of charge $q$. From \figurename{\ref{LbyOd}}, one can notice that the maximum instability shifts towards large values of $l$ for higher values of $r_{c}$. 

\section{Summary and Conclusions} 
In this article, we have analyzed the stability of timelike and null circular orbits in background of a 2+1 dimensional charged BTZ BH spacetime emerged in nonlinear electrodynamics model  using Lyapunov exponents as a tool and thus find the possibility of chaotic orbits. The main results are summarized below: 
\begin{itemize}
	\item[1.] (i) For massive test particles (timelike case), it is observed that the sign of the parameter $K$ determines the existence of timelike circular orbits. There are no circular orbits found for $K\geq 0 $ whereas for $K<0$, timelike circular orbits exist. We further derived the proper time Lyapunov exponent $\lambda_{p}$ and coordinate time Lyapunov exponent $\lambda_{c}$ in order to discuss the stability of circular orbits for charged BTZ BH. The timelike circular orbits are stable when the quantity $\Sigma>0$ i.e. the Lyapunov exponents become an imaginary quantity. 
	
	(ii) On the other hand, the circular orbits are unstable when $\Sigma<0$ i.e. the Lyapunov exponents are real. This instability in circular orbits indicates that chaos may be observed in geodesic motion i.e. some of these unstable circular orbits may indicate chaotic behavior. The orbits are marginally stable for $\Sigma=0$ i.e. the Lyapunov exponents vanish.
	
	\item[2.] For massless test particles (null case), the parameter ($K$) determines that the null circular orbits do not exist for $K\geq 0$ while they exist for $K<0$. There are two important parameters related to the instability of orbits, one is the coordinate time Lyapunov exponent $\lambda_{Null}$ and other is angular frequency $\Omega_{c}$. Both parameters are dependent of charge of BH ($q$) and parameter related to cosmological constant ($l$). The Lyapunov exponent $\lambda_{Null}$ is real and finite at radius of null circular orbits $r_{c}=-3K/4m$, which implies the existence of unstable null circular orbits, and this can be seen as a strong indication for occurrence of chaos in null circular orbits. 
	
	\item[3.] For the timelike case, we visually observed the behavior of Lyapunov exponents ratio  $-\lambda_{p}^{2}/\lambda_{c}^{2}$ as a function of radius of circular orbit $r_{0}$ and concluded that the ratio varies from one orbit to another orbit for different values of charge parameter $q$ and fixed value of mass $m$. 
	
	\item[4.] Moreover, we studied the behavior of the instability exponent $\lambda_{Null}/\Omega_{c}$ with respect to charge $q$ by varying the cosmological parameter $l$ and it is observed that the maximum instability shifted towards lower values of charge for large values of $l$. The variation of the same by varying radius of orbits $r_{c}$, the maximum instability shifted towards large values of $q$ for a large value of $r_{c}$. \\
	Beside this, the variation of the instability exponent with respect to cosmological parameter $l$ by varying charge parameter $q$ indicated that the maximum instability shifted towards lower values of $l$ for a large values of charge $q$, whereas by varying $r_{c}$, the maximum instability shifted towards the large values of $l$ for large values of $r_{c}$. 
	
	We intend to report on the issue of stability of various other BH spacetimes in GR and other alternative theories of gravity in near future in view of the techniques used in this paper.
\end{itemize}
\section*{\normalsize Acknowledgments}
{\normalsize One of the authors SG would like to thank the financial support provided by University Grants Commission (UGC), New Delhi, India as a Junior Research Fellow through UGC-Ref.No. {\bf 1479/CSIR-UGC NET-JUNE2017}. HN thankfully acknowledges Science and Engineering Research Board (SERB), India for financial support through grant no. {\bf EMR/2017/000339}. The authors also acknowledge the facilities available at ICARD, Gurukula Kangri (Deemed to be University) Haridwar those were used during the course of this work.  }  

\bibliographystyle{unsrt} 
\bibliography{shobitrefLYEX} 
\end{document}